Transformer For Low-frequency Extrapolating of Seismic Data

Zheng Cong, College of Instrumentation and Electrical Engineering, Jilin University, Changchun, China, 130026, cz19990714@126.com

Xintong Dong, College of Instrumentation and Electrical Engineering, Jilin University, Changchun, China, 130026, 18186829038@163.com

Shaoping Lu, School of Earth Sciences and Engineering, Sun Yat-Sen University, Guangzhou, China, Southern Marine Science and Engineering Guangdong Laboratory (Zhuhai), Zhuhai, China, and the Guangdong Provincial Key Lab of Geodynamics and Geohazards, Sun Yat-Sen University, Guangzhou, China, 510275, lushaoping@mail.sysu.edu.cn

Shiqi Dong, Department of Communication Engineering, Northeast Electric Power University, Jilin, China, 132012, dsq1994@126.com

Xunqian Tong, College of Instrumentation and Electrical Engineering, Jilin University, Changchun, China, 130026, txq@jlu.edu.cn

ABSTRACT

Full waveform inversion (FWI) is used to reconstruct the physical properties of subsurface media which plays an important role in seismic exploration. However, the precision of FWI is seriously affected by the absence or inaccuracy of low-frequency information. Therefore, reconstructing the low-frequency signals accurately is highly significant in seismic data processing. Low-frequency extrapolation of seismic records can be approached as a deep learning regression problem. Thus, to obtain low-frequency information from band-limited seismic records, a novel network structure called low-frequency extrapolation transformer (LFET) is proposed to construct the nonlinear mapping relationship between the data missing low-frequency and low-frequency data in a supervised learning approach, which is inspired by the transformer model widely used in natural language processing (NLP). We apply multi-head self-attention (MSA) modules to model the remote dependencies of seismic data. Based on this, we introduce a shifted window partitioning approach to reduce the calculating amount. Due to the field data are not suitable for supervised learning, we generate synthetic seismic records using submodels selected from the benchmark Marmousi model as training data whose characteristics are similar to that of the field data. A single trace of synthetic band-limited seismic data in the time domain is used as the input data, and the parameters of LFET are updated based on the errors between the predicted trace and the corresponding label. The experimental results on the data generated by different models, different wavelets, and different kinds of field marine data demonstrate the feasibility and generalization of the proposed method. Furthermore, the proposed method achieves higher accuracy with lower computational expense than the traditional CNN method.

# INTRODUCTION

Full waveform inversion (FWI) is a strong technique that reconstructs the subsurface properties with high accuracy (Chi et al., 2015; Guitton et al., 2020; Jiao et al., 2015) and can be used widely for structural imaging and comprehensive reservoir research (Mora, 1987; Lee and Kim, 2003; Warner and Guasch, 2016). The purpose of FWI is to obtain physical parameters from the complex seismic waves which are propagated in the subsurface media and collected by receivers distributed on the ground. The related parameters that can be calculated by FWI include longitudinal and shear waves, viscoelastic parameters, anisotropic parameters, and density (Guan and Tang, 1990; Kwon et al., 2015). The process of FWI is to make the synthetic data which is generated by an initial model based on prior information to fit the field data using all of the information through multiple iterations. However, FWI is a strongly nonlinear problem, low-frequency information in the data and an accurate initial model are keys to ensure the convergence and accuracy of the inverted results (Tarantola et al., 1986; Skarzynski et al., 2007). Low-frequency information in seismic data has a strong penetration ability, which can reduce the scattering and absorption of reflected waves during seismic wave propagation, thereby improving interpretation accuracy to the geological bodies with strong impedance. Cycle skipping during FWI is caused by the oscillatory nature of seismic data. Due to the synthetic seismic data is obtained based on a supposed velocity model, if the traveltime differences between waveforms in the same period of the synthetic and recorded data exceeds half a period, the FWI process will encounter cycle skipping (Xu et al., 2012; Xie et al., 2013). The low-frequency components which are insensitive to the traveltime differences because of their long wavelengths can provide an accurate large-scale velocity model to mitigate the cycle skipping problem (Wei, 2016; Virieux et al., 2017; Marjanović et al., 2019). Therefore, the low-frequency information contains large-scale underground changes which are important to prevent FWI from converging to the local minimum and essential to improving the accuracy of seismic inversion (Whitcombe et al., 2007; Virieux et al., 2009). However, due to the field data often contaminated by low-frequency noises and the limitation of acquisition equipment, the low-frequency information in received seismic data is usually muted or difficult to obtain directly.

In recent decades, some methods for extrapolating low-frequency data from the data missing low-frequency information have been proposed. The inversion scheme from the Laplace domain of data to generate an equivalent long-wavelength velocity model is proposed first (Shin and Cha, 2008). By using the ultra-low frequency (ULF) signal contained in the signal envelope, the velocity model with a large scale can be established (Wu et al., 2014). The waveform mode decomposition (WMD) method is proposed to reconstruct the low-frequency information of FWI with physics meaning to some extent (Hu et al., 2017). Zhang et al. (2017) used blind deconvolution to obtain the reflected impulse response which can be convolved with a designed low-frequency wavelet to obtain the corresponding low-frequency components. Li et al. (2018) acquired low-frequency components by using a nonlinear smoothing operator. Wang et al. (2018) moved the high frequency to the low part based on frequency shift theory. However, the low-frequency information extrapolated by these methods is often based on a nonlinear transformation of the data missing low frequencies, which may not achieve the desired accuracy or introduce extra noises.

Deep learning has been proven as a strong operator to achieve complex nonlinear mapping. With the development of computing device performance, deep learning methods are gradually applied in various fields (Hinton et al., 2006; LeCun et al., 2015; Guo et al., 2016; Kamilaris and Prenafeta-Boldú, 2018). For example, in the processing phase of seismic data: Cheng et al. (2022)

and Dong et al. (2023) apply deep learning methods to denoising complex seismic records. Zhong et al. (2023) realized the simultaneous denoising and reconstruction of seismic records in the processing phase of seismic data. Wang et al. (2023) use a transformer for inversion of the velocity model. Meanwhile, deep learning methods are gradually sovle many more complex problems in the field of seismic (Dong et al., 2022; Dong et al., 2024). In recent years, many low-frequency extrapolating methods based on deep learning have been proposed. Similarly, there are some methods proposed recently to achieve extrapolating low-frequency information accurately by artificial neural networks. Sun and Demanet (2020) constructed a training set based on the submodels of the Marmousi velocity model and designed a convolutional neural network (CNN) with 1-dimensional (1-D) convolutional blocks. By adopting a trace-to-trace training scheme, the effectiveness of the low-frequency signals predicted by the proposed method for the data obtained from Marmousi and BP 2004 benchmark models is proved. Luo et al. (2023) attempted a shot-based extrapolation and designed a multi-scale network to obtain features in different resolutions. Hu et al. (2019) introduced the beat tone information to the original phase data to predict low-frequency information from the input data.

In general, the quality of the training set and the structure of the network determines the fitting capability of the network. Here, we refer to the training scheme proposed by Sun and Demanet (2020) to feed the low-frequency-missed single trace in the time domain into a deep learning network to predict the corresponding low-frequency information. On one hand, the signals in the time domain provide a more detailed feature representation. On the other hand, although the 2-dimensional (2-D) shot gathers contain more features in the spatial domain, using the entire shot as input is not conducive to training and prediction due to the high computational cost. Besides, a sufficiently accurate 1-D prediction would also be spatially accurate.

In this study, we aim to incorporate the transformer model into the low-frequency extrapolation task as a data processing step instead of calculating the low-frequency gradient based on nonlinear transformation (Virieux and Operto, 2009; Zhou et al., 2012). Transformer is a novel model for building contextual relationships in sequences based on attention mechanisms, which is often used for natural language processing (NLP) tasks such as language modeling, language translation, text classification, and so on. Following are some reasons why we use a transformer: firstly, CNN models use convolutional layers to extract local features, but for long sequences like seismic traces, the receptive field needs to be enlarged by increasing the depth (Beltagy et al., 2020; Khan et al., 2022). As a deep learning model based on multi-head self-attention (MSA) blocks (Rao et al., 2021; Zhai et al., 2022), the transformer delivers greater efficiency by capturing long-distance dependencies through global modeling (Wang et al., 2018; Dosovitskiy et al., 2020). Secondly, the lightweight research aimed at transformers has made exciting progress (Li et al., 2021; Meng et al., 2022). By applying the window partitioning approach to self-attention, the training time and prediction speed of a transformer can be close to that of the CNNs (Liang et al., 2021; Liu et al., 2021). Finally, the transformer model with a larger depth allows it to fit more complex functions with greater accuracy and generalization (Han et al.,2022; Zhang et al., 2022).

In this work, we obtain submodels from the Marmousi velocity model (Bourgeois et al., 1991), and synthetic data sets are obtained by waveform equation modeling based on the acoustic media. In general, deep learning methods are represented by combining many different functions to approximate ideal complex nonlinear relationships. For the low-frequency data prediction task, our purpose is to build a mapping relationship between low-frequency missing seismic data with low-frequency seismic data, while the global modeling capabilities of the transformer make it more suitable for processing long sequences like traces. Besides, there are strong correlations between the waveforms (amplitude, phase) of band-limited data and low-frequency data in the time domain (Hu et al., 2017) Therefore, we adopt a transformer to complete the low-frequency data prediction task. On this basis, we apply window-based multi-head self-attention (WMSA) to the transformer to save computational costs and reduce the time required for prediction. The transformer is trained on a synthetic dataset to learn the mapping relationship between low-frequency data and low-frequency missing data. We proved the precision and availability of extrapolated low-frequency information to FWI on the SEG/EAGE salt dome model. Besides, the influence of different wavelets and noise is also considered. The proposed method is compared to traditional CNN to prove the feasibility and generalization by experiments on different benchmark models and field marine data. In addition, the proposed method achieves accurate predictions with higher computational efficiency.

## METHODS

### Network Architecture

*A. Multi-head self-attention*

As the core idea of transformers, the self-attention mechanism has a unique advantage in dealing with sequences by considering the relations between each element and the other elements. Considering the sequence properties of seismic traces in the time domain which indicates the wave propagation in the subsurface media, the application of self-attention is a viable method for mapping the features between the events in low-frequency information and the band-limited data along the whole time axis. The self-attention mechanism calculates the correlation weights of each element in the sequence with all others. These weights indicate the relationships between the elements, which can better obtain the context information of a seismic trace in a long time-axis by modeling the remote dependencies. To be specific, as shown in Figure 1, self-attention is achieved in the following ways:

$$(\mathbf{Q}, \mathbf{K}, \mathbf{V}) = \text{input} \times (W^{\mathbf{Q}}, W^{\mathbf{K}}, W^{\mathbf{V}}),$$

$$\text{Self} - \text{attention}(\mathbf{Q}, \mathbf{K}, \mathbf{V}) = \text{softmax}(\mathbf{Q}\mathbf{K}^{\mathrm{T}})\mathbf{V}, \quad (1)$$

where $W^{\mathbf{Q}}, W^{\mathbf{K}}, W^{\mathbf{V}}$ are trainable matrices, $\mathbf{Q}, \mathbf{K}, \mathbf{V}$ means linear transformation of the input obtained by multiplying by $W^{\mathbf{Q}}, W^{\mathbf{K}}, W^{\mathbf{V}}$, $\mathbf{K}^{\mathrm{T}}$ means the transpose matrix of $\mathbf{K}$. We first computed the inner product of $\mathbf{Q}$ and $\mathbf{K}^{\mathrm{T}}$ to obtain the correlation coefficient matrix of the data. The correlation coefficient matrix contains the correlation index of each two elements. After being processed by the softmax function, the correlation indexes are normalized to [0,1] and assigned to the linear transformation result $\mathbf{V}$ to finish the weight updating. From the basis, the MSA mechanism achieves to focus on different aspects of information by dividing the input into multiple subspaces and calculating them separately, thus improving the representation ability of the model, such as:

$$\text{Multi} - \text{head self} - \text{attention}(\mathbf{Q}, \mathbf{K}, \mathbf{V}) = \text{Concat}(\text{head}_1, \dots, \text{head}_i, \dots, \text{head}_h),$$

$$\text{head}_i = \text{softmax}(\mathbf{Q}_i \mathbf{K}_i^{\mathrm{T}})\mathbf{V}_i, \quad (2)$$

where $i$ means the serial number of attention heads, and the total number of attention heads is h. Here, we split $\mathbf{Q}, \mathbf{K}, \mathbf{V}$ h heads on the channel dimension respectively. After calculating the self-attention of each head, the calculation of all attention heads is combined to get the output.

*B. Low-Frequency Extrapolation Transformer*

As shown in Figure 2(a), a transformer block (TB) consists of an MSA module, a multi-layer perceptron (MLP) module, and two layer-norm (LN) layers. MLP module is a feedforward neural network, which contains an input layer, a hidden layer, and an output layer. The input layer expands the number of channels of input data by four times through linear transformation. We apply an activation function called Gaussian Error Linear Unit (GeLU) into the hidden layer. Then, the output layer recovers the number of channels of data through a linear transformation. MLP introduces nonlinear characteristics to the model, thus enhancing the expressiveness of the model (Tolstikhin et al., 2021). LN layers normalize the input to ensure its mean is 0 and its variance is 1, thus improving the stability and training effect of the model. (Ba et al., 2016).

Seismic data usually have a long time length, which leads to a large amount of computation. Besides, the down-sampling operation will lose the information of traces, thus destroying the frequency distribution of the trace. Aimed at this problem, we applied window-based transformer blocks (WTB) and shifted window-based transformer blocks (SWTB) to low-frequency prediction tasks (depicted in Figures 2(b) and 2(c)). The operation in the WTB can be expressed as:

$$X_a = \text{WMSA}(\text{LN}(X_{in})) + X_{in,} \tag{3}$$

$$X_b = \text{MLP}(\text{LN}(X_a)) + X_a,$$

where $X_{in}$ represents the input of WTB, $X_a$ and $X_b$ represents the output of WMSA and MLP. In addition, skip connections are adopted to improve the generalization ability of the network.

As shown in Figure 2(e), WTB divides the trace into several non-overlapping windows, thus achieving to reduce the complexity of computing by limiting the calculation of MSA to each window, separately. Based on this, SWTBs are adopted to promote the information interaction between adjacent windows by moving the windows. To be specific, our network structure is summarized in Figure 2(d). Firstly, the input is processed by a convolutional layer to expand the channel from 1 to

64 to obtain more features for learning. Then, these features are treated alternately by WTB and SWTB 4 times in total. Noteworthy, the number of heads is set to 8, the size is set to 40, and the shift step is set to 20. Finally, the convolutional layer recovers 64 channels into 1 to generate the overall network prediction.

*C. Low-Frequency Extrapolation Theory*

In this study, we trained the network to establish a mapping relationship between low-frequency data and the data missing low-frequency information:

$$\hat{L} = M(A, \theta), \qquad (4)$$

where $\hat{L}$ means predicted low-frequency data by the network, $A$ means the data missing low-frequency entering the network, $M$ represents the mapping relationship built by the transformer, and $\theta$ represents the parameters of the network. Then, the loss function is the mean square error (MSE) which is as follows:

$$\text{loss}(\theta) = \frac{1}{2N} \lVert M(A_j, \theta) - L_j \rVert, \qquad (5)$$

where $N$ means batch size, $A_j$ and $L_j$ represent the input of the network and the labeled data, $j$ represents the serial number of the input and the labeled data. The parameters of the network are updated by the Adam algorithm, and the updating of parameters is finished when the loss does not change significantly.

*D. Construction of Training Datasets*

Transformer models usually require a large amount of data for training. However, low-frequency information in field data with enough accuracy is rare which makes it hard to support supervised learning. Thus, producing plenty of synthetic data that is similar to the field data to generate training datasets is necessary. In this part, seventeen submodels were extracted from the

Marmousi velocity model to build the training datasets. Figure 3 shows the selection of 9 submodels based on evenly location, and 8 complex submodels selected manually. The shape of each submodel measures 32×96 and is interpolated to 128×384 with a grid spacing of 20m. Additionally, a water layer with a depth of 100m is added on top. The training datasets were simulated using MATLAB, detailed parameters are shown in Table 1. For the input, a Ricker wavelet with a dominant frequency of 7Hz is processed by a 10th-order Butterworth high-pass filter with a cut-off frequency of 5Hz was used as the source signal. The label was processed by a 10th-order Butterworth low-pass filter with a cut-off frequency of 5Hz after being excited by a Ricker wavelet with a dominant frequency of 7Hz. The sampling rate and the recording time are set to 0.002s and 2.4s. To collect training datasets, 96 shots were evenly excited from 0m to 7680m on the surface of the water layer. Additionally, 384 receivers were placed evenly from 0m to 7680m. The high amplitude of the direct wave is not favorable for the convergence of training. To mute the direct wave, we adopted the same arrangement of shots and receivers on a water model which has the same size as the marmousi model to obtain the pure direct wave. The difference between synthetic models and direct waves is the data for training(Sun and Demanet, 2020).

After seismic modeling, a total number of 17×96×384 traces are obtained. The traces adjacent to each other are similar. Therefore, we sample the data with a sampling step of 2 to reduce the duplication of data to improve training efficiency. 313344 pairs of traces with a length of 2400 are obtained as the training dataset. Because data normalization can promote network convergence, we adopted the normalization method proposed by Jin et al (2021) which amplifies weak signals. The formulation and inverse operator are as follows:

$$T(x) = sign(x)log_{10}(|ax| + 1) \qquad (6)$$

$$T^{-1}(y) = a^{-1}sign(y)(10^{|y|} - 1) \tag{7}$$

where $x$ and $y$ represent the trace and its transformed result, and $a$ is a parameter to control the level of boosting. Before the training of the network, both the input data and label data are operated by $T(\cdot)$. For the test case, seismic records are first divided into traces, and each trace is transformed and then input into the network. Finally, these predicted results are then combined into seismic records after inverse transformation.

*E. Experimental Environment and Training Parameters*

For the deep learning method, hardware devices have a significant impact on the effectiveness of training and processing. In this study, our experimental environment is composed of a CPU (Intel Xeon Silver 4210R Processor 13.75M Cache 2.40 GHz), and NVidia GeForce RTX 3090 (24 GB RAM). All training and testing are mainly finished based on pytorch (1.10.2). As shown in Table 2, the batch size is set to 32. The initial learning rate is set to $10^{-3}$ and drops to 0.1 times for every 20 training epochs, the total of training epochs is 60.

NUMERICAL EXAMPLES

This section demonstrates the performance of extrapolated FWI with LFET in four parts. Notably, we use U-Net as a CNN method for comparison, which is also performed with the same training dataset. The parameters of LFET and U-Net are detailed in Table 3. In the first part, we present the capability of LFET and U-Net to predict low frequency (0-5Hz) from band-limited data (above 5Hz) on the Marmousi model. In the second part, we compared the generalization of the two methods with the salt dome model. Additionally, we leverage the low-frequency data predicted by LFET for the salt dome model to perform FWI, validating the utility of the processing result for

FWI applications. Finally, we applied LFET and U-Net to marine seismic streamer recordings and ocean bottom cable (OBC) field recordings to verify the effectiveness of the method in actual situations.

*Low-Frequency Extrapolation on the Marmousi Model*

To verify the accuracy of extrapolated low frequency, we generate synthetic records based on a full-size Marmousi velocity model for testing. Figure 4(a) shows the comparison of the 50th trace of processing results in the time domain. The processing results of LFET and U-Net are both close to the ground truth, which demonstrates the feasibility of deep learning methods for low-frequency extrapolation tasks. Nevertheless, Figures 4 (b) and (c) provide detailed observations of the high amplitude area between 0.4 and 1.2s and the weak signal area between 1.2 and 2.4s, respectively (highlighted in Figures 4(a) by red boxes). As shown in Figure 4(b), the amplitudes of some early arrivals reconstructed by U-Net do not match enough to the ground truth. In addition, the late arrivals that are predicted by U-Net have obvious errors (Figure 4(c)). Compared to U-Net, the results predicted by LFET are smoother and more in line with the labeled data. Figure 4(d) compares the results of the two methods in the frequency domain. There is almost no information in the frequency band of 0-2 Hz band in the input high-frequency component (black line), thus limiting the processing accuracy of FWI. It can be observed that U-Net has errors in low-frequency bands. In contrast, the LFET fits the frequency spectrum of the ground truth extremely well, which proves that the LFET is substantially more feasible and more accurate. In summary, we compared the predictions of two methods applied to synthetic records with missing low frequencies in the Marmousi velocity model in the time and frequency domain. The result demonstrates that LFET has strong learning ability and achieves accuracy beyond the conventional CNN-based methods.

*Generalization Ability Testing*

In this part, a SEG/EAGE salt dome model is adopted to verify the generalization ability of the proposed method in the case of different models. Different from the Marmousi model, there is a salt body in the middle of the velocity model (marked in the red box), which increases the probability of falling into local minima in the inversion (Figure 5).

Figure 6 illustrates the processing results of single-shot in the time domain. Figures 6(a) and 6(b) depict the high-frequency component (above 5Hz) and low-frequency component (0-5Hz) of synthetic salt dome model data, which are the input and labeled data for the network. In Figure 6(c), we observe the predicted low frequency by U-Net, while Figure 6(d) presents the predicted low frequency using LFET. U-Net demonstrates proficiency in reconstructing early arrivals, but the late arrivals that have low amplitude are erroneous (indicated by red arrows). In contrast, LFET produces low-frequency data with a clearer and more continuous representation of weak signals, aligning more closely with the true low-frequency data. For the aspect of amplitude, LFET preserves it better and maintains the same values as the true low-frequency component.

The analysis of results in the f-k domain for both methods is presented in Figure 7. In Figure 7(a), the input data contains little information below 5Hz, which is unsuitable for direct application in FWI. Figure 7(b) illustrates the f-k spectrum for the true low-frequency data (0-5 Hz). As shown in Figure 7(c), U-Net recovers an approximate frequency band of the low-frequency information. However, artifacts are discernible in its f-k spectrum. In contrast, LFET (Figure 7(d)) exhibits more accurate predictions of low-frequency details, aligning closely with the true low-frequency component. Compared to U-Net, LFET demonstrates superior accuracy and amplitude preservation. In summary, we utilize a SEG/EAGE salt dome model to assess the generalization of LFET for different models. The analysis consists of both the time domain and f-k domain for the results of a single shot. The results of LFET fit the true data better, demonstrating that LFET has more powerful modeling capabilities and that generalization for different models, thus obtaining meaningful low-frequency components from the high-frequency component of the data.

To assess the availability of the reconstructed low-frequency information, we employ the predicted low-frequency results obtained from LFET in FWI. Figure 8(a) shows the resulting model starting from the linear initial model, which is far from the true velocity model (Figure 5). Subsequently, Figure 8(b) shows the results of FWI starting from the linear initial model with the data above 5Hz. It can be observed that due to the absence of low-frequency data, the inversion results have an obvious cycle-skipping phenomenon, resulting in the missing of a large amount of physical meaning of the information. Figure 8(c) displays the resulting model starting from the linear initial model and utilizing the low-frequency data extrapolated by LFET. This model effectively captures the low-wavenumber information of the salt dome, illustrating the reliability of low-frequency information obtained by LFET. However, it still has a slight cycle-skipping phenomenon. Finally, we used the high-frequency data for inversion which starts from the low-wavenumber model constructed from the low frequencies extrapolated by LFET. As depicted in Figure 8(d), the resulting model is clearer and more matched with the true velocity model, demonstrating that the cycle-skipping phenomenon has been alleviated. In summary, we adopt the low-frequency data reconstructed by LFET to FWI to verify the availability of LFET. The results show that LFET can provide accurate low-frequency information to build a suitable low-wavenumber velocity model to relieve the cycle skipping phenomenon in FWI.

***Low-Frequency Extrapolation on Field Records***

In this part, we assess the practicality of the proposed method by applying it to field records. To be specific, a field marine streamer record and an OBC field record are processed (Figures 9(a) and 10(a)). Compared to synthetic records, field records have obvious differences and are more complex, thus increasing the difficulty of prediction.

We used a shot of field marine streamer data to test the practical application of our method. The sampling interval is 2 ms, and the sampling length is 4800 points. The shot consists of 150 traces with a trace spacing of 12.5 m. Different from our training dataset, the dominant frequency of the field marine seismic streamer data is 16 Hz. Therefore, we filtered the field records to obtain more suitable data for model processing. Specifically, the band-limited data (5-8 Hz) is used as input. Figure 9(b) shows the low-frequency (0-5Hz) component of the field marine streamer data, which contains some continuous events and valid information with physical meaning. Figure 9(c) shows the prediction of U-Net. Although U-Net can predict early arrivals effectively, the part about late arrivals is blurry and the amplitude of the weak signal is much different from the actual low-frequency component (indicated by the green arrows). Figure 9(d) shows the processing result of LFET. A prominent promotion has been made to the low-frequency component, such as more balanced energy, and more consistent amplitude. It is worth noting that the proposed method achieves more continuous events than the true low-frequency component in some regions (marked by the yellow boxes), which demonstrates the application significance of the proposed method in field data.

Furthermore, we adopted OBC field data to validate the generality of the proposed method. The sampling interval is 2 ms and the sampling length is 4800 points. The shot consists of 220 traces with a trace spacing of 12.5 m. The low-frequency (0-5Hz) component of the OBC field data is shown in Figure 10(b). Due to equipment limitations, the low-frequency component lacked meaningful information, leading to a cycle-skipping phenomenon. Similar to the approach with marine streamer data, the band-limited OBC data (5-8Hz) is fed into the network. The processing result of U-Net is shown in Figure. 10(c), the low-frequency information at early arrivals lacks continuity and the energy is unbalanced at late arrivals with tiny amplitude. Conversely, LFET (depicted in Figure 10(d)) exhibits a more continuous representation of early arrivals and late arrivals are clearer. To sum up, these results demonstrate the feasibility of LFET for the low-frequency extrapolation of field data. Meanwhile, our method exhibits superior effectiveness and accuracy compared to the traditional CNN method.

## DISCUSSION

In this section, we further analyze the generalization of the proposed method based on more complex cases. Firstly, we construct a Marmousi model excited by a Ricker wavelet with a phase rotation to investigate the performance of the proposed method under different wavelet conditions. Secondly, we analyze the impact of noise on the prediction results. Thirdly, we compare the computational and efficiency of LFET and U-Net. Meanwhile, structure similarity index measure (SSIM) and root mean square error (RMSE) are used as quantitative indicators of the performance. Finally, we research the effect of transformer parameters on both efficiency and effectiveness of the proposed method.

### *Condition of Phase Rotation*

To check the extrapolation capability of the LFET in the context of data excited by other wavelets, we train the LFET with the aforementioned training set but we test it with a phase-rotated wavelet. To be specific, we employed a seismic source signal by rotating the phase of the Ricker wavelet at a 135° angle to excite the Marmousi model for testing. Figures 11(a) and 11(b) depict the band-limited (above 5Hz) data and true low-frequency data, respectively. Figure 11(c) illustrates that the result of U-Net roughly matches the true data, but it still has limitations in restoring weak signals (indicated by the green arrows), which reduces the accuracy of FWI. In contrast, LFET is more fit to the true low-frequency data (Figure 11(d)). In particular, the weak signal reconstructed by LFET is closer to the true data and clearer (indicated by the yellow arrows), which shows that LFET achieves higher accuracy. To sum up, deep learning methods still perform well in the case of different wavelets demonstrating their generalization. In addition, LFET exhibits an accuracy advantage in reconstructing weak signals, thereby affirming its superior modeling capabilities.

*Robustness analysis*

In this section, we evaluate the robustness of our proposed network using the Marmousi model in the presence of additive noise. To quantify the impact of noise on the test, we introduced 5% Gaussian noise to the synthetic records to construct the noisy training sets. Subsequently, the noisy records are filtered (5-15Hz) and normalized to generate the noise-containing training set. Figures 12(a), 12(b), and 12(c) present the original data, noisy data, and true low-frequency data for testing, respectively. It is evident that the introduced noise significantly affects the ability of the network to extract meaningful information. As depicted in Figure 12(d), the prediction of U-Net still exhibits noticeable noise, which is unfavorable for the identification of weak signals in late arrivals. Conversely, LFET performs effectively even in the presence of noise interference and obtains superior results compared to U-Net (Figure 12(e)). In conclusion, the robustness of deep learning methods against noise interference can be enhanced by introducing noise to the training set. Even though noise hurts the accuracy of extrapolated low frequencies to some extent, the prediction of deep-learning methods has a degree of robustness. Moreover, LFET exhibits superior anti-noise capabilities compared to U-Net when both models are trained on the same dataset, which emphasizes the effectiveness and robustness of LFET in processing noisy conditions.

*Computational Efficiency and Quantitative Comparisons*

For deep learning methods, assessing computational cost is crucial for determining feasibility. Table 3 provides a comparative analysis of the training time and processing time for LFET and U-Net. Meanwhile, SSIM and RMSE of single-shot processing are also considered quantitative indices. SSIM and RMSE are defined by the following formulas:

$$RMSE = \sqrt{\frac{1}{XY}\sum_{m=1}^{X}\sum_{n=1}^{Y}(l(m,n) - p(m,n))}, \quad (8)$$

$$SSIM = \frac{(2\mu_l\mu_p + C1)(2\sigma_{lp} + C2)}{(\mu_l^2 + \mu_p^2 + C1)(\sigma_l^2 + \sigma_p^2 + C2)}, \quad (9)$$

where $X$ and $Y$ represent trace number and time samples of the seismic record. $l(m,n)$ and $p(m,n)$ denote the $(m,n)$th point of labeled data and prediction, respectively. In formula (8), $l$ and $p$ present a single shot of the true low-frequency component and processing results, $\mu_\bullet$ and $\sigma_\bullet$ present the mean and standard deviation of •, $\sigma_{lp}$ presents the covariance between $l$ and $p$. $C1$ and $C2$ are constants used to maintain stability. And $C1 = (k_1 L)^2$ $C2 = (k_2 L)^2$, where $k_1 = 0.01$, $k_2 = 0.03$, $L=255$ presents the dynamic range of the images. In general, the RMSE reflects the disparity between the predicted result and the true value, while SSIM reflects the degree of similarity between the aforementioned pair. Specifically, when RMSE approaches 0 or SSIM reaches 1, it indicates a higher level of accuracy in the predictive model.

To eliminate the influence of anomalous data in the test, we selected 50 records containing near-offset and far-offset for testing. The final index was determined by calculating the average value. Table 4 displays the training and processing times for LFET, which amount to 3.16h and 18.83s, respectively. In comparison, U-Net requires longer training and processing times, registering at 3.54h and 23.37s, respectively. It is noteworthy that both the training time and processing time of LFET are shorter than those of U-Net, reflecting the computational efficiency advantages of LFET over U-Net. Meanwhile, the SSIM and RMSE of LFET are 0.8912 and 0.0146, respectively. In contrast, the SSIM and RMSE of U-Net are 0.5897 and 0.1217, respectively, falling short of the corresponding indexes for LFET. In summary, the LFET achieves higher processing accuracy in comparison to traditional CNN methods. It is worth noting that benefits from its utilization of a window-based attention approach, LFET achieves a heightened accuracy while maintaining higher computational efficiency. In summary, we prove that LFET not only achieves superior processing accuracy in contrast to traditional CNN methods but also demonstrates enhanced computational efficiency, leveraging the advantages of the window-based attention approach.

*Module Availability Analysis*

In this part, we investigate the impact of LFET parameters, including the number of transformer modules, attention heads, and window sizes, on computational cost and effectiveness. Table 5 presents the relevant computational parameters and evaluation indicators. For the number of transformer modules, the reduction of them will decrease the complexity of LFET, thus reducing the prediction accuracy. Conversely, an increase in transformer modules introduces more parameters, leading to greater training difficulties and decreased prediction accuracy. As shown in Table 5, the processing time and training time of LFET with 4 transformer layers are 12.67s and 1.43h, while the 20-layer configuration requires 32.69s and 6.38h, respectively. The SSIM values for 4 and 20 layers are 0.8894 and 0.8753, with corresponding RMSE values of 0.0193 and 0.0179, indicating that more or fewer layers are not effective in improving accuracy. Therefore, it is crucial to select the appropriate number of layers for the deep-learning method to avoid both excessive and insufficient amounts. For the number of attention heads, it has a significant effect on training time and a minor effect on processing time. Multiple attention heads in MSA contribute to increased computational complexity during training, resulting in longer training times. Meanwhile, more attention heads may make the information too scattered for the network to pick up attention information, while less attention heads may limit the network to obtain various information. As illustrated in Table 5, LFET with 4 attention heads shows processing and training times of 15.40s and 2.28h, whereas the configuration with 16 attention heads requires 15.66s and 4.43h. The SSIM values for 4 and 16 attention heads are 0.8253 and 0.8896, with corresponding RMSE values of 0.0234 and 0.0197. Notably, both more and fewer attention heads than 8 adversely affect prediction accuracy, which proves the importance of fine-tuning this parameter. Finally, the processing time and training time are both affected by the size of the window. WMSA and SWMSA compute self-attention within local windows to reduce the parameters of the network, thus decreasing the time required for training and processing. Meanwhile, too large attention windows may extract unnecessary information, while a window that is too small may limit the network's learning ability. Here, LFET with a window size of 10 exhibits processing and training times of 15.23s and 3.16h, whereas the window size of 80 requires 17.65s and 3.41h, which proves that the larger window size results in more computation. Meanwhile, the SSIMs of LFET with window sizes of 10 and 80 are 0.8776 and 0.8725, while their RMSEs are 0.0186 and 0.0190, respectively. Both of them have lower accuracy than the LFET with the window size of 40, which proves that the window size of 40 is an appropriate parameter for LFET. In summary, our thorough analysis examines the impact of various network parameters on computational cost and performance, emphasizing the importance of parameter selection. Through experimentation, we identified the optimal combination of parameters as 8 layers, 8 attention heads, and a window size of 40, thus obtaining the highest prediction accuracy.

## CONCLUSION

Deep learning is applied to extrapolate low-frequency information to overcome the cycle-skipping problem in FWI. In this study, we propose a transformer-based network named LFET which precisely extracts global features from band-limited data to predict the associated low-frequency data. The training datasets for LFET are built by seismic modeling on the submodels selected from the Marmousi velocity model. The trained model shows its generalization on the data generated on the salt-dome model and generated by a phase-rotated wavelet. Meanwhile, the trained LFET is also robust in processing noisy records to some extent by retraining the model with noisy training datasets. The results of FWI show that the high-precision prediction of LFET effectively alleviates the cycle-skipping problem. The successful application of the trained LFET on the field marine data and OBC field data demonstrates the strong generalization of the proposed method. Besides, LFET has faster processing speeds and shorter training times due to the window-based attention approach. The design of LFET based on a transformer provides novel ideas for low-frequency extrapolation of seismic data, which are typical sequences in the time domain.

# REFERENCES

Ba, J. L., Kiros, J. R. and Hinton, G. E, 2016, Layer normalization: arxiv preprint arxiv:1607.06450.

Beltagy, I., Peters, M. E. and Cohan, A, 2020, Longformer: The long-document transformer: arxiv preprint arxiv:2004.05150.

Bourgeois, A., Bourget, M., Lailly, P., Poulet, M., Ricarte, P., Versteeg, R., and Grau, G, 1991, Marmousi, model and data: The Marmousi Experience, 5-16.

Cheng, M., Lu, S. P., Dong, X. T., Zhong, T, 2022, Multiscale recurrent-guided denoising network for distributed acoustic sensing-vertical seismic profile background noise attenuation: Geophysics, **88**, no 1, 201-217.

Chi, B., Dong, L., Liu, Y, 2015, Correlation-based reflection full-waveform inversion: Geophysics, **80**, no. 4, 189-202.

Dong, X. T., Lin, J., Lu, S. P., Huang, X. G., Wang, H. Z., Li, Y, 2022, Seismic Shot Gather Denoising by Using Supervised-Deep-Learning Method with Weak Dependence on Real Noise Data: a Solution to the Lack of Real Noise Data: Surveys in Geophysics, **43**, no. 5, 1363-1394.

Dong, X. T., Lu, S. P., Cong, Z., Zhong, T, 2023, Multi-stage Residual Network for Intense DAS Background Noise Attenuation: Geophysics, **88**, no. 6, 181-198.

Dong, X. T., Lu, S. P., Lin, J., Zhang, S. K., Ren, K., Cheng, M, 2024, Can Deep-Learning Compensate the Sparse Shots in Imaging Domain? A Potential Alternative for Reducing the Acquisition-Cost of Seismic Data: Geophysics, **89**, no. 2, 1-87.

Dosovitskiy, A., Beyer, L., Kolesnikov, A., Weissenborn, D., Zhai, X.H., Unterthiner, T., Dehghani, M., Minderer, M., Heigold, G., Gelly, S., Uszkoreit, J., Houlsby, N, 2020, An image is worth 16x16 words: Transformers for image recognition at scale: arxiv preprint arxiv:2010.11929.

Guan, L. P. and Tang, Q. J, 1990, High/low frequency compensation of seismic signal: Geophysical Prospecting for Petroleum, **29**, no 3, 35-45.

Guitton, A., Ayeni, G., Díaz, E., 2012, Constrained full-waveform inversion by model reparameterization: Geophysics, **77**, no. 2: 117-127.

Guo, Y., Liu, Y., Oerlemans, A., Lao, S., Wu, S. and Lew, M. S, 2016, Deep learning for visual understanding: A review: Neurocomputing, **187**, 27-48.

Han, K., Wang, Y., Chen, H., Chen, X., Guo, J., Liu, Z., Tang, Y., Xiao, A., Xu, C., Xu, Y., Yang, Z., Zhang, Y. and Tao, D, 2022, A survey on vision transformer: IEEE transactions on pattern analysis and machine intelligence, **45**, no. 1: 87-110.

Han, L. G., Zhang, Y., Han, L. and Yu, Q. L, 2012, Compressed sensing and sparse inversion based low-frequency information compensation of seismic data: Journal of Jilin University (Earth Science Edition), **42**, 259-264.

Hinton, G. E., Osindero, S. and Teh, Y. W, 2006, A fast learning algorithm for deep belief nets: Neural computation, **18**, no. 7, 1527-1554.

Hu, W., Jin, Y. C., Wu X.Q., Chen J.F, 2019, A progressive deep transfer learning approach to cycle-skipping mitigation in FWI: SEG International Exposition and Annual Meeting, SEG-2019-3216030.

Hu, Y., Han, L., Xu, Z., Zhang, F. and Zeng, J, 2017, Adaptive multi-step full waveform inversion based on waveform mode decomposition: Journal of Applied Geophysics, **139**, 195-210.

Jiao, K., Sun, D., Cheng, X. and Vigh, D, 2015, Adjustive full waveform inversion: SEG International Exposition and Annual Meeting, SEG-2015.

Jin, Y. C., Hu, W., Wang, S. R., Zi, Y., Wu, X. Q. and Chen, J. F, 2021, Efficient progressive transfer

learning for full-waveform inversion with extrapolated low-frequency reflection seismic data: IEEE Transactions on Geoscience and Remote Sensing, **60**, 1-10.

Kamilaris, A. and Prenafeta-Boldú, F. X, 2018, Deep learning in agriculture: A survey: Computers and electronics in agriculture, **147**, 70-90.

Khan, S., Naseer, M., Hayat, M., Zamir, S. W., Khan, F. S. and Shah, M, 2022, Transformers in vision: A survey: ACM computing surveys (CSUR), **54**, 1-41.

Kwon, T., Seol, S. J. and Byun, J, 2015, Efficient full-waveform inversion with normalized plane-wave data: Geophysical Journal International, **201,** no 1, 53-60.

LeCun, Y., Bengio, Y. and Hinton, G, 2015, Deep learning: nature, **521**, no 7553, 436-444.

Lee, K. H. and Kim, H. J, 2003, Source-independent full-waveform inversion of seismic data: Geophysics, **68**, no 6, 2010-2015.

Li, Y. Y. and Demanet, L, 2016, Full-waveform inversion with extrapolated low-frequency data: Geophysics, **81**, no 6, 339-348.

Liang, J., Cao, J., Sun, G., Zhang, K., Van Gool, L. and Timofte, R, 2021, Swinir: Image restoration using swin transformer: In Proceedings of the IEEE/CVF international conference on computer vision, 1833-1844.

Liu, Z., Lin, Y., Cao, Y., Hu, H., Wei, Y., Zhang, Z., Lin, S. and Guo, B, 2021, Swin transformer: Hierarchical vision transformer using shifted windows. In Proceedings of the IEEE/CVF international conference on computer vision, 10012-10022.

Luo, R., Gao, J. and Meng, C, 2023, Low-Frequency Prediction Based on Multiscale and Cross-Scale Deep Networks in Full-Waveform Inversion: IEEE Transactions on Geoscience and Remote Sensing, **61**, 1-11.

Marjanović, M., Plessix, R. É., Stopin, A. and Singh, S. C, 2019, Elastic versus acoustic 3-D full waveform inversion at the East Pacific Rise 9° 50′ N: Geophysical Journal International, **216**, no 3, 1497-1506.

Mora, P, 1987, Nonlinear two-dimensional elastic inversion of multioffset seismic data: Geophysics, **52**, no. 9, 1211-1228.

Mulder, W. A. and Plessix, R. E, 2008, Exploring some issues in acoustic full waveform inversion: Geophysical Prospecting, **56**, no 6, 827-841.

Rao, R. M., Liu, J., Verkuil, R., Meier, J., Canny, J., Abbeel, P., Sercu, T. and Rives, A, 2021, MSA transformer: International Conference on Machine Learning, **139**, 8844-8856.

Shin, C., and Cha, Y. H., 2008, Waveform inversion in the Laplace domain: Geophysical Journal International, **173**, no. 3, 922-931.

Skarzynski, H., Lorens, A., Piotrowska, A. and Anderson, I, 2007, Preservation of low frequency hearing in partial deafness cochlear implantation (PDCI) using the round window surgical approach: Acta oto-laryngologica, **127**, no. 1, 41-48.

Sun, H. and Demanet, L, 2020, Extrapolated full-waveform inversion with deep learning: Geophysics, **85**, no. 3, 275-288.

Sun, H. and Demanet, L, 2021, Deep learning for low-frequency extrapolation of multicomponent data in elastic FWI: IEEE Transactions on Geoscience and Remote Sensing, **60**, 1-11.

Tarantola, A., 1986, A strategy for nonlinear elastic inversion of seismic reflection data: Geophysics, **51,** no. 10, 1893-1903.

Tolstikhin, I. O., Houlsby, N., Kolesnikov, A., Beyer, L., Zhai, X., Unterthiner, T., Yung, J., Steiner, A., Keysers, D., Uszkoreit, J., Lucic, M., and Dosovitskiy, A, 2021, Mlp-mixer: An all-mlp architecture for vision: Advances in neural information processing systems, 34, 24261-24272.

Virieux, J. and Operto, S, 2009, An overview of full-waveform inversion in exploration geophysics: Geophysics, **74**, no 6, 1-26.

Virieux, J., Asnaashari, A., Brossier, R., Métivier, L., Ribodetti, A. and Zhou, W, 2017, An introduction to full waveform inversion: Encyclopedia of exploration geophysics. Society of Exploration Geophysicists, 1-1.

Wang, H. Z., Lin, J., Dong, X. T., Lu, S. P., Li, Y., Yang, B. J, 2023, Seismic velocity inversion transformer: Geophysics, 2023, **88**, no 4, 513-533.

Wang, X., Girshick, R., Gupta, A. and He, K, 2018, Non-local neural networks: Proceedings of the IEEE conference on computer vision and pattern recognition,7794-7803.

Warner, M. and Guasch, L, 2016, Adaptive waveform inversion: Theory: Geophysics, **81**, no. 6, 429-445.

Wei, J. D, 2016, Geophone deconvolution low-frequency compensation for seismic data: Oil Geophysical Prospecting, **51**, no. 2, 224-231.

Whitcombe, D., and Hodgson, L, 2007, Stabilizing the low frequencies: The Leading Edge, **26**, no. 1, 66-72.

Wu, R. S., Luo, J. and Wu, B, 2014, Seismic envelope inversion and modulation signal model: Geophysics, **79**, no. 3, 13-24.

Xie, X. B, 2013, Recover certain low-frequency information for full wave-form inversion: SEG Annual Meeting, 1053–1057.

Xu, S., Wang, D., Chen, F., Zhang, Y. and Lambare, G, 2012, Full waveform inversion for reflected seismic data: 74th EAGE Conference and Exhibition incorporating EUROPEC 2012. European

Association of Geoscientists & Engineers, cp-293-00729.

Zhai, X., Kolesnikov, A., Houlsby, N. and Beyer, L, 2022, Scaling vision transformers: Proceedings of the IEEE/CVF Conference on Computer Vision and Pattern Recognition, 12104-12113.

Zhang, Z., Zhang, H., Zhao, L., Chen, T., Arik, S., Pfister, T, 2022, Nested hierarchical transformer: Towards accurate, data-efficient and interpretable visual understanding: Proceedings of the AAAI Conference on Artificial Intelligence, **36**, no 3, 3417-3425.

Zhong, T., Cong, Z., Wang, H., Lu, S., Dong, X., Dong, S. and Cheng, M, 2023, Multi-Scale Encoder-Decoder Network for DAS Data Simultaneous Denoising and Reconstruction: IEEE Transactions on Geoscience and Remote Sensing, **61**, 1-15.

Zhou, H., Amundsen, L. and Zhang, G, 2012, Fundamental issues in full waveform inversion: SEG International Exposition and Annual Meeting, SEG-2012.

LIST OF FIGURES

Figure 9. Extrapolation results on marine seismic streamer data. (a) band-limited (above 5Hz) data, (b) low-frequency component of the marine seismic streamer data (0–5 Hz), (c) prediction of U-Net, (d) prediction of LFET.

Figure 10. Extrapolation results on OBC field data. (a) band-limited (above 5Hz) data, (b) low-frequency component of the OBC field data (0–5 Hz), (c) prediction of U-Net, (d) prediction of LFET.

Figure 11. Extrapolation results on phase-rotated Marmousi model. (a) band-limited (above 5Hz) data, (b) true low-frequency data (0–5 Hz), (c) prediction of U-Net, (d) prediction of LFET.

Figure 12. Discussion of noise robustness. (a) band-limited (above 5Hz) data, (b) noisy band-limited (above 5Hz) data, (c) true low-frequency data (0–5 Hz), (d) prediction of U-Net, (e) prediction of LFET.

# LIST OF TABLES

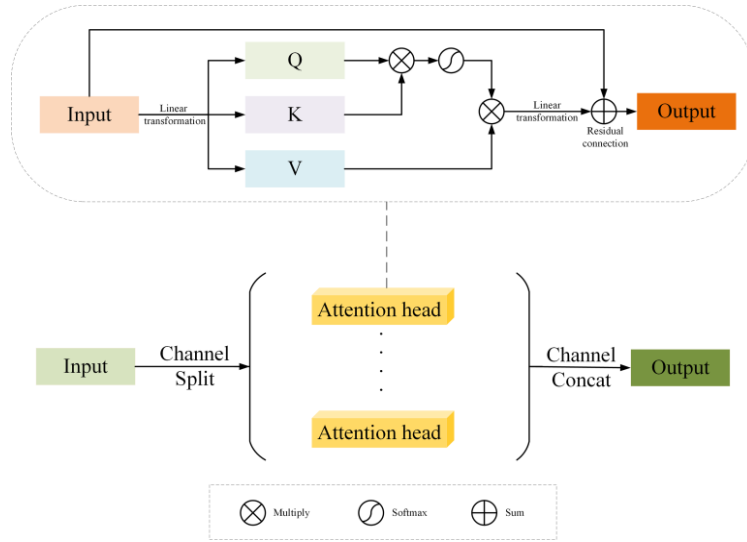

Figure 1. Structure of MSA.

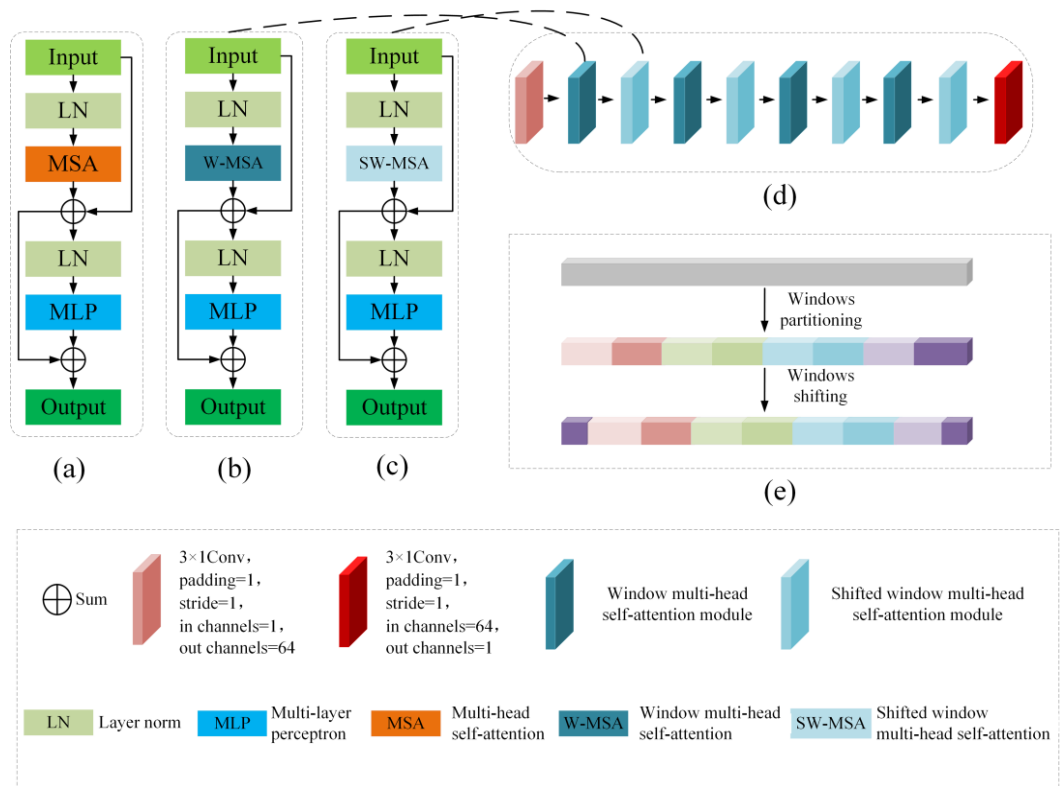

Figure 2. Network structure of the proposed method. (a) The structure of TB. (b) The structure of WTB. (c) The structure of SWTB. (d) The structure of LFET.

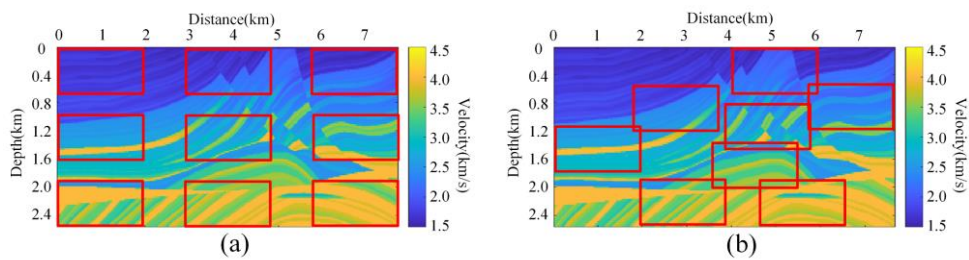

Figure 3. The Marmousi velocity model used for collecting the training data sets. (a) 9 submodels

selected on location. (b) 8 complex submodels.

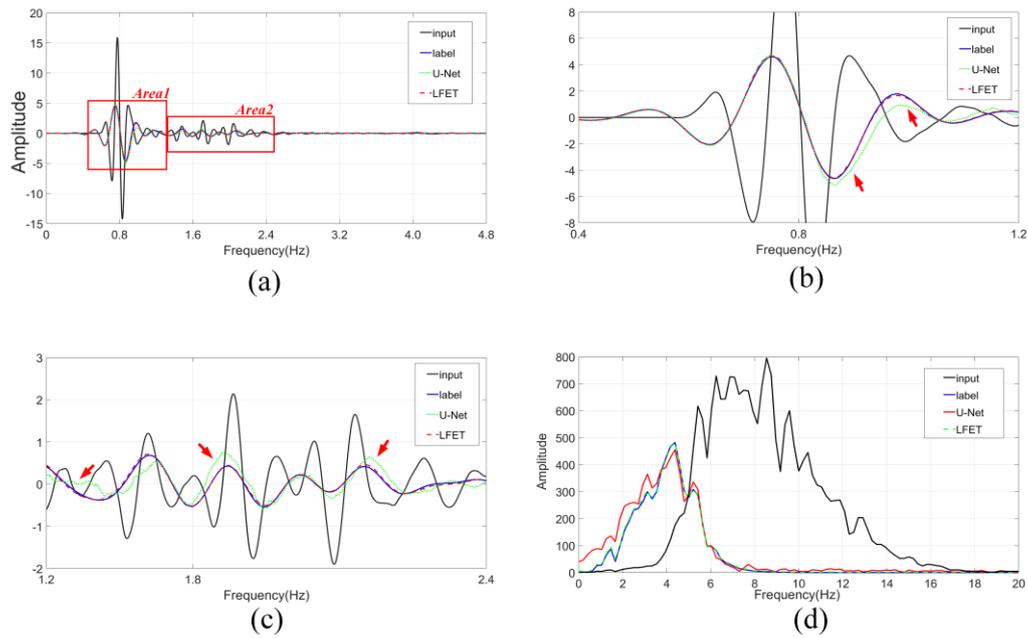

Figure 4. Comparison of 50th trace of processing results. (a) 50th trace in the time domain. (b) 50th trace in the time domain of 0.4-1.2s. (c) 50th trace in the time domain of 1.2-2.4s. (d) 50th trace in frequency domain.

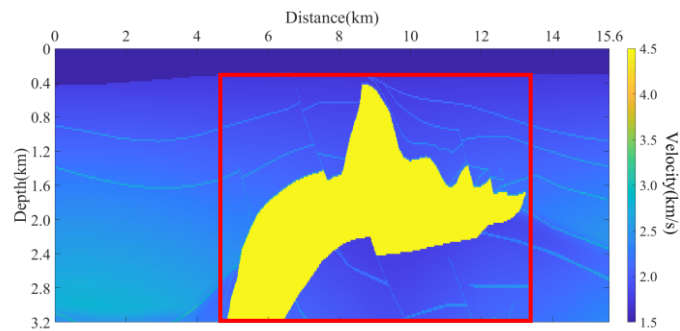

Figure 5. The salt dome model. (The salt geobody is marked in the red box)

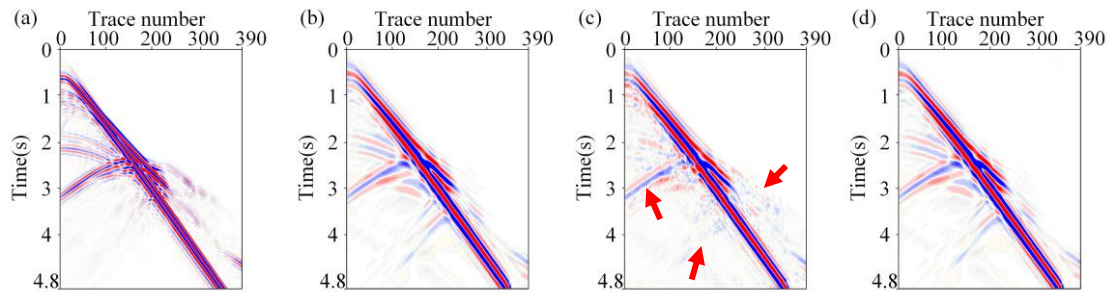

Figure 6. Extrapolation results on synthetic salt dome model. (a) band-limited (above 5Hz) recordings, (b) true low-frequency recordings (0–5 Hz), (c) prediction of U-Net, (d) prediction of LFET.

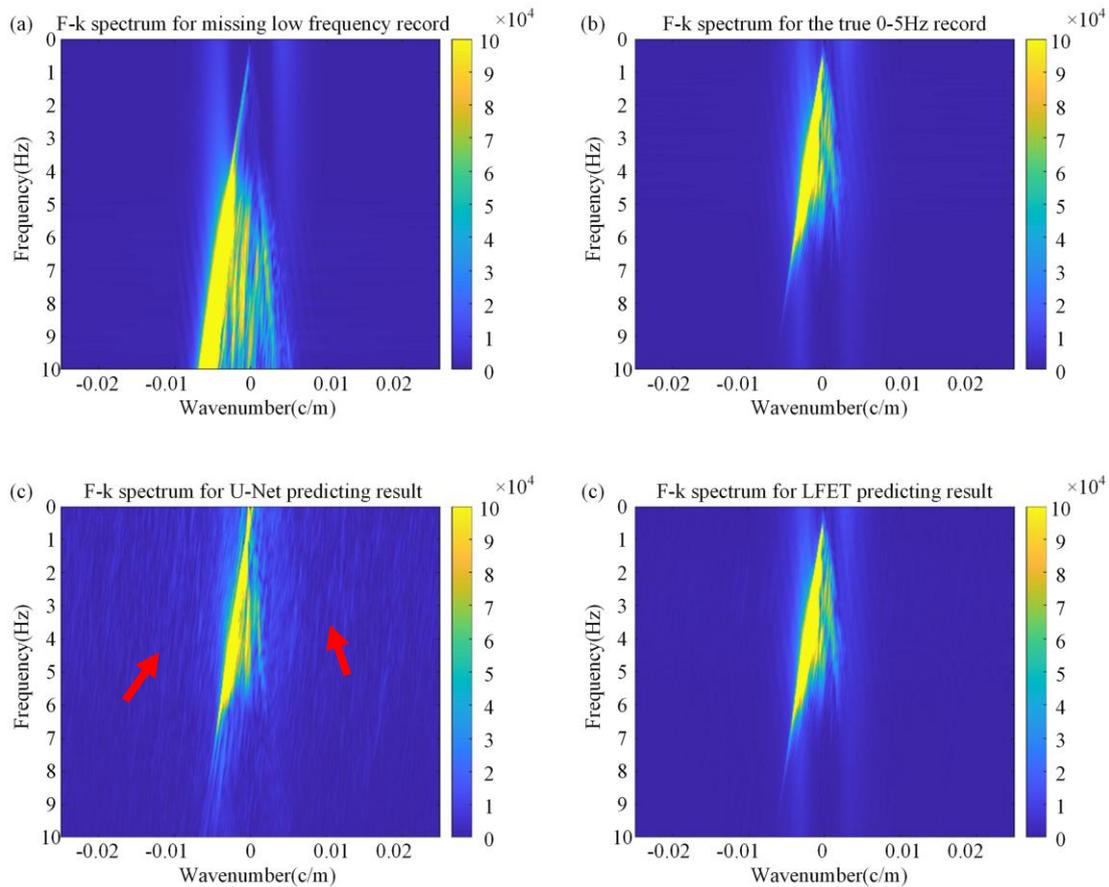

Figure 7. Extrapolation results on synthetic salt dome model in f-k domain. (a) f-k spectrum for band-limited (above 5Hz) recordings, (b) f-k spectrum for true low-frequency recordings (0–5 Hz), (c) f-k spectrum for prediction of U-Net, (d) f-k spectrum for prediction of LFET.

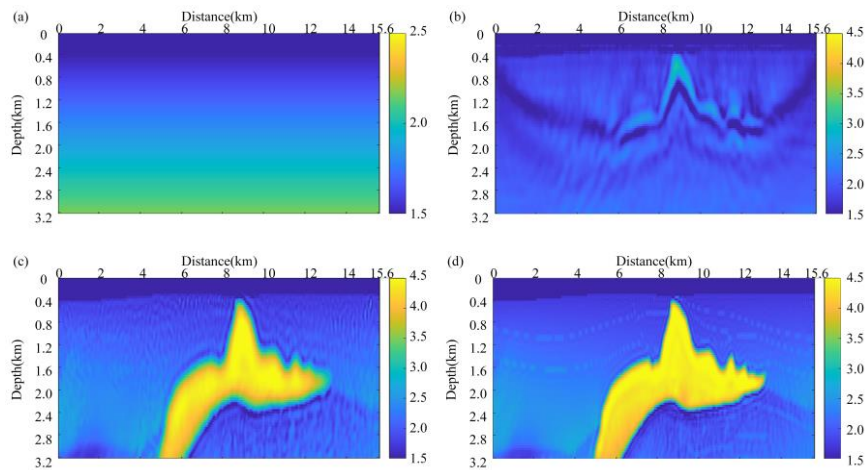

Figure 8. Comparison of the inverted models from FWI. (a) The resulting model starts from the original initial model. (b) The resulting model starts from the high-frequency data (above 5Hz). (c) The resulting model starts from the extrapolated data below 5 Hz which is recovered by the high-frequency data (above 5Hz). (d) The resulting model starts from the low-wavenumber velocity

model constructed with the low frequencies extrapolated by LFET.

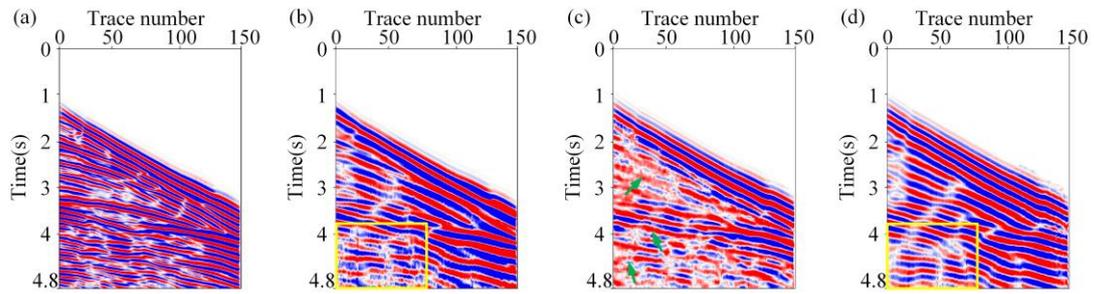

Figure 9. Extrapolation results on marine seismic streamer data. (a) band-limited (above 5Hz) data, (b) low-frequency component of the marine seismic streamer data (0–5 Hz), (c) prediction of U-Net, (d) prediction of LFET.

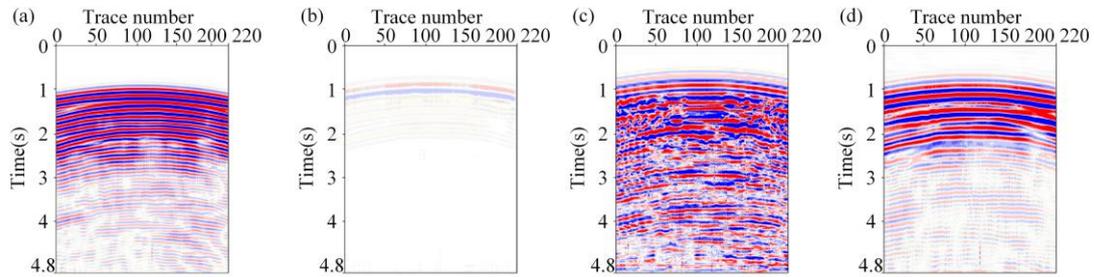

Figure 10. Extrapolation results on OBC field data. (a) band-limited (above 5Hz) data, (b) low-frequency component of the OBC field data (0–5 Hz), (c) prediction of U-Net, (d) prediction of LFET.

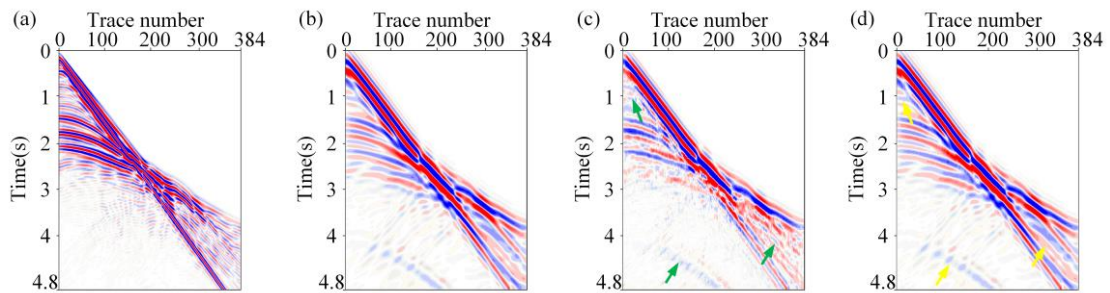

Figure 11. Extrapolation results on phase-rotated Marmousi model. (a) band-limited (above 5Hz) data, (b) true low-frequency data (0–5 Hz), (c) prediction of U-Net, (d) prediction of LFET.

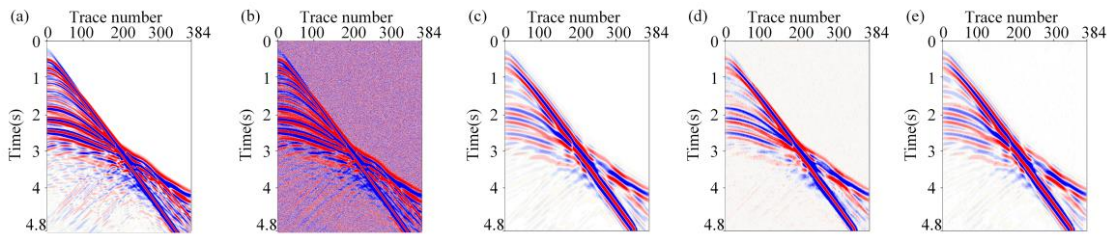

Figure 12. Discussion of noise robustness. (a) band-limited (above 5Hz) data, (b) noisy band-limited

(above 5Hz) data, (c) true low-frequency data (0–5 Hz), (d) prediction of U-Net, (e) prediction of LFET.

Table 1. Parameters of the forward modeling method.

| Parameters | Setting |
|---|---|
| Seismic wavelet | Ricker |
| Dominant frequency | 7Hz |
| Trace interval | 20m |
| Grid space | 20m |
| Sampling interval | 0.002s |
| Total recording time | 4.8s |
| Wave velocity | 1500-4000m/s |
| Wave equation | Acoustic |

Table 2. Training parameters of LFET.

| Parameters | LFET |
|---|---|
| Optimizer | Adam |
| Trace length | 2400 |
| Batch size | 32 |
| Epoch number | 60 |
| Learning rate range | $[10^{-3},10^{-5}]$ |
| Learning rate change interval | 20 |

Table 3. Network parameters of LFET and U-Net.

| Parameters | LFET | U-Net |
|---|---|---|
| Convolution layer | 2 | 27 |
| Transformer block | 8 | 0 |
| Number of attention heads | 8 | - |
| Window size | 40 | - |
| Shift step | 20 | - |

Table 4. Computational efficiency and quantitative comparisons.

|  | Original record U-Net | LFET |
|---|---|---|
| Processing time (s) | 23.37 | 15.60 |
| Training time (hour) | 3.54 | 3.17 |
| RMSE | 0.1217 | 0.0146 |
| SSIM | 0.5897 | 0.8912 |

Table 5. Quantitative data analysis of different methods.

| Module usage | Processing time (s) | Training time (h) | RMSE | SSIM |
|---|---|---|---|---|
| LFET (8 transformer layers) | 15.60 | 3.17 | 0.0146 | 0.8912 |
| LFET (4 transformer layers) | 12.67 | 1.43 | 0.0193 | 0.8894 |
| LFET (20 transformer layers) | 32.69 | 6.38 | 0.0179 | 0.8753 |
| LFET (4 attention heads) | 15.40 | 2.28 | 0.0234 | 0.8253 |
| LFET (16 attention heads) | 15.66 | 4.43 | 0.0197 | 0.8896 |
| LFET (window size=10) | 15.23 | 3.16 | 0.0186 | 0.8776 |
| LFET (window size=80) | 17.65 | 3.41 | 0.0190 | 0.8725 |